\documentclass[aps,preprint,showpacs,amsmath,amssymb,epsfig]{revtex4}
\usepackage{graphicx}% Include figure files
\usepackage{bm}% bold math

\begin{document}

\title{Implications of Graviton-Graviton Interaction to Dark Matter.}

\author{
A.~Deur$^{\nuva}$\footnote{Present address: Thomas Jefferson National 
Accelerator Facility, Newport News, VA23606, USA}.}

\affiliation{
\baselineskip 2 pt
\centerline{{$^{\nuva}$University of Virginia, Charlottesville, VA 22904}}}

\newcommand{\nuva}{1}

\date{\today}

\begin{abstract}
Our present understanding of the universe requires
the existence of dark matter and dark energy. We describe here
a natural mechanism that could make exotic dark matter and possibly dark
energy unnecessary. Graviton-graviton interactions increase the 
gravitational binding of matter. This increase, for large massive systems 
such as galaxies, may be large enough to make exotic dark matter superfluous. 
Within a weak field approximation we compute the effect on the rotation curves
of galaxies and find the correct magnitude and distribution 
without need for arbitrary parameters or additional exotic particles. 
The Tully-Fisher relation also emerges naturally from this framework.
The computations are further applied to galaxy clusters. 

\end{abstract}


\maketitle

%%%%%%%%%%%%%%%%%%%%%%%%%%%%%%%%%%%%%%%%%%%%%%%%%%%%%%%%%%%%%%%%%%%%%%%
Cosmological observations appear to require ingredients
beyond standard fundamental physics, such as exotic dark 
matter~\cite{Dark Matter} and dark energy~\cite{review dark energy}. In this 
Letter, we discuss whether the observations suggesting the existence of dark matter 
and dark energy could stem from the fact that the carriers of gravity, 
the gravitons, interact with each others. In this Letter, we will
call the effects of such interactions {}``\emph{non-Abelian}''.
The discussion parallels similar phenomena in particle physics and so we will use 
this terminology, rather than the one of General Relativity, although we believe
it can be similarly discussed in the context of General Relativity. We will
connect the two points of view wherever it is useful.

Although massless, the gravitons interact with each other because of the 
mass-energy equivalence. The gravitational coupling $G$ is very small so
one expects $G^2$ corrections to the Newtonian potential due to graviton-graviton  
interactions to be small in general. However, gravity always attracts
(gravitons are spin even) and systems of large mass $M$ can produce intense
fields, balancing the smallness of $G$. Indeed, $G^2$ corrections
have been long observed for the sun gravity field since they induce 
the precession of the perihelion of Mercury. Such effects 
are calculable for relatively weak fields, using either the Einstein field 
equations (the non-linearity of the equations is related to the non-Abelian nature 
of gravity~\cite{Feynman gravity}), or Feynman graphs in which the one-graviton 
exchange graphs produce the Newtonian (Abelian) potential and higher order 
graphs give some of the $G^2$ corrections. Gravity self-coupling must be included 
in these 
calculations to explain the measured precession~\cite{Feynman gravity}. The 
non-Abelian effects increase gravity's strength which, if large enough, would mimic 
either extra mass (dark matter) or gravity law modifications such as the
 empirical MOND model~\cite{MOND}.

Galaxies are weak gravity field systems with stars moving at non-relativistic speeds. 
For weak fields, the Einstein-Hilbert action can be rigorously expanded in a power 
series of the coupling $k$ ($k^2 \propto G$) by developing the metric $g_{\mu \nu}$ 
around the flat metric $\eta_{\mu\nu}$. This is known (see e.g. refs.~\cite{Salam},\cite{Kiefer}) 
but we recall it for convenience: $g_{\mu\nu}$ is parametrized, e.g. 
$g_{\mu\nu}=\left(e^{k \psi}\right)_{\mu\nu}$, and expended around $\eta_{\mu\nu}$. It leads to: 
\begin{eqnarray}
\frac{1}{16 \pi G}{\int d^{4}x\sqrt{-g}g_{\mu\nu}R^{\mu\nu} = \int d^{4}x(\partial\psi\partial\psi+} 
k \psi\partial\psi\partial\psi+k^2 \psi^{2}\partial\psi\partial\psi+...)+k \psi_{\mu \nu} T^{\mu \nu}\label{eq:Einstein-Hilbert} \end{eqnarray}
\noindent Here, $g=det~g_{\mu \nu}$, $R^{\mu\nu}$ is the Ricci tensor,
$\psi^{\mu\nu}$ is the gravity field, and $T^{\mu \nu}$ is the source (stress-energy) tensor. 
Since our interest is $\psi$ self-interactions, we will not include the 
source term in the action. (We note that it does
not mean that $T^{\mu \nu}$ is negligible: we will use later the fact that 
$T^{00}$ is large. It means that $T^{\mu \nu}$ is not a relevant degrees of freedom
in our specific case. This will be further justified later.)  
A shorthand notation is used for the terms $\psi^n\partial\psi\partial\psi$ 
which are linear combinations of terms having this form for which the Lorentz indexes 
are placed differently. For example, the explicit form of the shorthand
$\partial\psi\partial\psi$ is given by the Fierz-Pauli Lagrangian~\cite{Fierz-Pauli} 
for linearized gravity field.  

\begin{figure}[ht!]
\begin{center}
\centerline{\includegraphics[scale=0.2, angle=0]{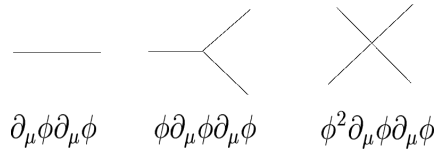}}
\end{center}
\vspace{-1.7cm}
\caption{Graviton interactions corresponding to the terms of the Lagrangian, 
Eq.~\ref{eq:Lagrangian}.}
\label{fig:fg}
\end{figure}

The Lagrangian $\mathcal{L}$ is a sum of 
$\psi^{n}\partial\psi\partial\psi$. These terms can be transformed 
into $\frac{1}{n+1}$$\psi^{n+1}$$\partial^{2}\psi$ by integrating by part in 
the action $\int d^{4}x\mathcal{L}$. We consider first the $\partial\psi\partial\psi$
term. The Euler-Lagrange equation of motion obtained by varying
the Fierz-Pauli Lagrangian leads to $\partial^{2}\psi^{\mu\nu}=-k^2 
(T^{\mu\nu}-\frac{1}{2}\eta^{\mu\nu}Tr(T))$.
Since the $T^{00}$ component dominates $T^{\mu\nu}$ within the stationary weak
field approximation, so too $\partial^{2}\psi^{00}$ dominates 
$\partial^{2}\psi^{\mu\nu}$ and one can keep only the $\psi^{00}$ terms in 
$\psi\partial^{2}\psi$, i.e. in $\partial\psi\partial\psi$. Finally, after applying 
the harmonic gauge condition $\partial^{\mu}\psi_{\mu\nu}=\frac{1}
{2}\partial_{\nu}\psi_{\kappa}^{\kappa}$,
we obtain for the first term in $\mathcal{L}$ $\partial\psi\partial\psi\rightarrow\frac{1}
{4}\partial_{\lambda}\psi^{00}\partial^{\lambda}\psi_{00}$. Higher order terms 
proceed similarly since they are all of the form 
$\frac{1}{n+1}\psi^{n+1}\partial^{2}\psi$~\cite{higher order}.
The factor in front of each $\psi^{n}\partial\psi\partial\psi$ ($n \neq 0$) may 
depend however on how 
$g_{\mu\nu}$ is expanded around $\eta_{\mu \nu} $. For this reason, and because
the higher order terms are complicated to derive, we use a different
approach to determine the rest of the Lagrangian: we build it from the appropriate 
Feynman graphs (see Fig.~\ref{fig:fg}) 
using, with hindsight of previous discussion, 
only the $\psi^{00}\equiv\phi$ component of the field. Each term in the Lagrangian 
corresponds to a Feynman graph: the terms quadratic, cubic and quartic in $\phi$  
correspond respectively to the free propagator, 
the three legs contact interaction and the four legs contact interactions. The forms 
$\phi\partial\phi\partial\phi$ and $\phi^{2}\partial\phi\partial\phi$
(rather than $\phi^{3}$ or $\phi^{2}\partial\phi$ for example for the three legs 
graph) are imposed by the dimension of $G$. (Note that in
Eq.~\ref{eq:Einstein-Hilbert}, the origin of the two derivatives in the generic 
form $\phi^{n}\partial\phi\partial\phi$
is from the two derivatives in the Ricci tensor and the absence of
derivative in $g_{\mu\nu}$).

Since we are considering the total field from all
particles, then (neglecting here non-linear effects and binding energies) 
$k^2=\sum m \ 16 \pi G$ where $m$ is the
nucleon mass and $\sum m=M$ with $M$ the total mass of the 
system. This may be modeled with a space that is discretized with a lattice spacing 
$d$. We are interested in
the attraction between two cubes of $d^{3}$ volume filled with the 
gravity fields generated by $N$ sources of similar masses and homogeneous distribution. 
Since we are unable to treat $N$ sources we consider only a global field. 
Under the field superposition principle, the magnitude 
of the total field in each cube is proportional to $N$. 
As gravity always attracts, we used a global coupling $\sum_{1}^{N}mG=MG$~\cite{m+m}.
 
Under these simplifications and hypotheses we obtain from 
Eq.~\ref{eq:Einstein-Hilbert}~\cite{factorials}:
\begin{eqnarray}
{\int d^{4}x\mathcal{L}=\sum_{\mu=1}^{4}\int d^{4}x(\partial_{\mu}\phi\partial_{\mu}\phi+}
\frac{\sqrt{16\pi GM}}{3!}\phi\partial_{\mu}\phi\partial_{\mu}\phi+\frac{16\pi GM}{4!}\phi^{2}\partial_{\mu}\phi\partial_{\mu}\phi+...).\label{eq:Lagrangian} \end{eqnarray}
\noindent

To quantify gravity's non-Abelian effects on galaxies, we have used 
numerical lattice techniques: A Monte Carlo Metropolis algorithm 
was employed to estimate the two-point correlation function (Green function) 
that gives the potential. To test our Monte Carlo, we computed 
the case for which the high-order terms of $\mathcal{L}$ are set 
to zero, and recovered the expected Newtonian potential or, when a 
fictitious mass $m_{\phi}$ is assigned to the field $\phi$, the expected 
Yukawa potential $V(r)\propto{(e}^{-m_{\phi}r})/r$. We also insured the 
independence of our results from the lattice size
and the physical system size. In our calculations, the usual 
circular boundary conditions cannot be used. A pathological example is
the one of a linear potential, for which a 
simulation with such conditions would return  an irrelevant 
constant rather than the potential. Instead of circular 
boundary conditions, we set the boundary nodes of the lattice to be 
random with an average zero value. These nodes were never updated. 
In addition, although we update the fields on the nodes close to 
the boundary nodes, we did not use them in the calculation of the 
Green function (in the results presented, we ignored the 4 
nodes closest to the lattice boundary. We varied this number and 
found compatible results).

\begin{figure}[ht!]
\begin{center}
\vspace{-0.7cm}
\centerline{\includegraphics[scale=0.27, angle=0]{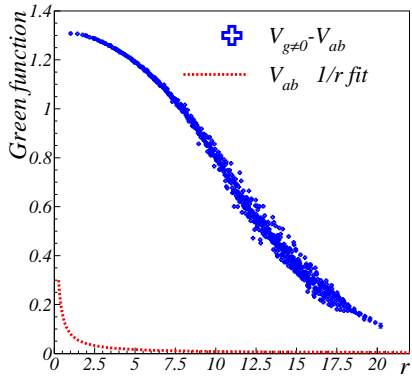}}
\end{center}
\vspace{-1.5cm}
\caption{Two-point Green function, after subtracting the 
$1/r$ contribution (dotted line). $r$ is in lattice units.}
\label{fig:lattice results}
\end{figure}

The Green function is shown in Fig.~\ref{fig:lattice results}. 
These results are for a lattice of size $L=28d$, with $d$ the 
lattice spacing, $\sqrt{\frac{16\pi GM}{d}}=4.9\times10^{-3}$ 
($d$ converts the coupling to lattice units), $M=10^{10}$ 
M$_{\odot}$ and $d=1$ kpc. For $r\gtrsim5$, the Green function 
is roughly linear. 

We now apply our calculations to the case of galaxies. 
For homogeneous distributions with spherical symmetry, the net field distortion 
from the force carrier self-interaction cancels out~\cite{geometry}. 
Similarly, a cylindrical symmetry reduces the effect. 
To first order we treat a spiral galaxy as a thin disk with a cylindrical symmetry 
and approximate $V(r)$ in Fig.~\ref{fig:lattice results}
as linear: the force is of constant value $b$. We are interested
in the force between the disk center and the circumference points.
The field lines point evenly outward, so the force at any point on the circumference
is reduced by 2$\pi r$: the force is then $b/(2\pi r)$ and $V(r)=b~ln(r)/(2\pi)$.
Adding back the (unaffected) {}``Abelian'' part $a/r$, we obtain:

\vspace{-0.5cm}
\begin{eqnarray}
V(r)=-GM\left(\frac{1}{r}+\frac{b}{2\pi a}ln(r)\right)\label{eq:pot disc 2}\end{eqnarray}
(For a homogeneous spherical distribution, the constant force 
becomes $b/4\pi r^{2}$, leading to a Newtonian potential 
$\left(a+b/4\pi\right)/r\propto1/r$, as expected. We note that typically, 
$a \gg b/4\pi$.)

\begin{figure}[ht!]
\begin{center}
\centerline{\includegraphics[scale=0.48, angle=0]{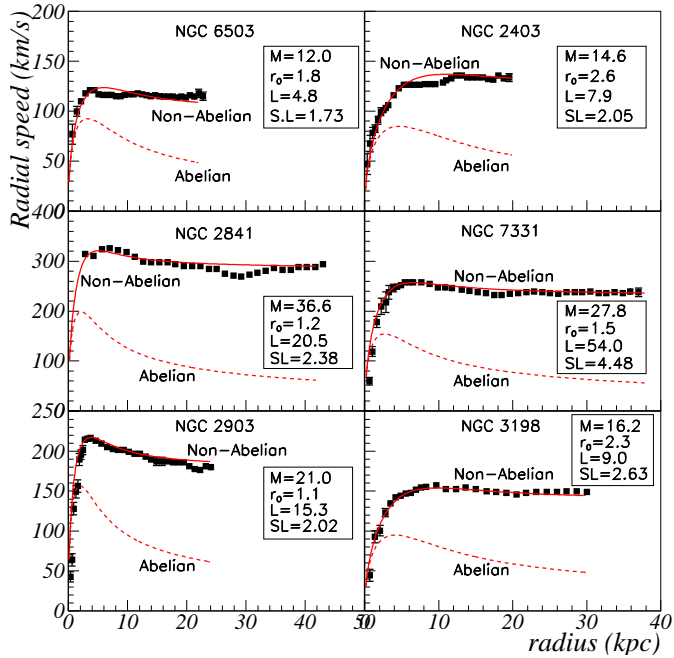}}
\end{center}
\vspace{-1.5cm}
\caption{Computed rotation curves (continuous
lines) compared to the measurements (squares) and the curves without field self-interaction (dashed lines). We used the values of R$_{25}$
given in Ref.~\cite{data rot curve} for the galaxy radii, and the values given on 
each plots for the parameters $M$ and $r_{0}$. 
The luminosity $L$ of the galaxies and the values of $r_{0}$ 
(noted scale length, SL) from Ref.~\cite{data rot curve} are also given for comparison
(units are $10^{9}$ M$_{\odot}$ for $M$ and $L$ and kpc for $r_{0}$).}
\label{fig:all rot curves}
\end{figure}

We can now look at rotation curves for spiral galaxies. Those, shown
in Fig.~\ref{fig:all rot curves}, are obtained by calculating $a$
and $b$ for given galaxy masses and sizes and assuming an exponential
decrease of the galaxy density with its radius: $\rho(r)=\frac{M}{2\pi r_{0}^{2}}e^{-r/r_{0}}$.
Galaxy luminous masses and sizes being not well known, we adjusted
$M$ and $r_{0}$ to best fit the data. They can be compared to the luminosity $L$ 
of the galaxies and the values of $r_{SL}$ from
Ref.~\cite{data rot curve} also given in Fig.~\ref{fig:all rot curves}
We did not use $L$ in the simulation but report it since it indicates
a lower bound for $M$ (consequently, NGC7331 for which $M<L$ pauses 
a problem within our simple spiral galaxy model). 
The curves reproduce well the data given our simple model of galaxy.
In addition to our rough approximation in of modeling 
galaxies, it should be emphasized that while our results should conservatively 
be viewed as indicating quantitatively the self-coupling 
effects~\cite{note non-Abelian} of the 
gravity field, there are several caveats: 1) The particular choice of 
boundary conditions may generate a non-physical artifact, although we checked
within the means of our lattice simulation that this was not the case; 
2) There are approximations inherent to a lattice 
calculation, in particular the cut-off on the high energy modes due to the 
lattice finite spacing; 3) Approximations are used to go 
from the Einstein-Hilbert action to the polynomial scalar action; 
4) We have used an approximate magnitude for the field self-coupling of $\sqrt{GM}$, 
which neglects non-linear effects and the specific distribution of sources in 
the studied system.

\begin{figure}[ht!]
\begin{center}
  \centerline{\includegraphics[scale=1.2, angle=0]{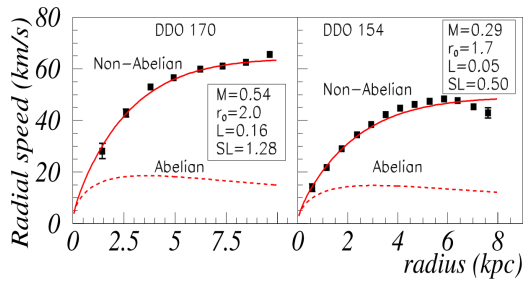}}
\end{center}
\vspace{-1.5cm}
\caption{Dwarf galaxy rotation curves.}
\label{fig:rot curve dwarves}
\end{figure}

The calculation applies similarly to dwarf galaxies. Results for 
galaxies DDO 170 and DDO 153 are shown in Figure
~\ref{fig:rot curve dwarves}.
The results agree with the observation that the luminous mass together
with a Newtonian potential contributes especially little to dwarf galaxy
rotation curves. 

The Tully-Fisher relation~\cite{TF}  is readily explainable by non-Abelian effects. 
This empirical law relates linearly the log of a galaxy mass to the log of 
its rotation velocity $v$: $ln(M)=\alpha ln(v)-\beta$, with $\alpha=3.9\pm0.2$ 
and $\beta\simeq1.5$. Equating the centripetal force acting on a object of mass $m$, 
$-mv^2\overrightarrow{u_r}/r$ ($\overrightarrow{u_r}$ is the unit vector) to the 
gravitational force 
$-GMmb\overrightarrow{u_r}/(2\pi ar)$ given by the potential of 
Eq.~\ref{eq:pot disc 2} for large distances yields ${v^2}=GMb/(2\pi a)$. 
Since $a$, the coefficient of the Newtonian potential in Eq.~\ref{eq:pot disc 2} 
is proportional to GM ($a=\tau GM$) and since $b$ can only be a function of the field 
self-coupling magnitude, $b(\sqrt{GM})$, we have 
${v^2}=b(\sqrt{GM})/(2\pi \tau)$, which correlates $v$ and $M$ since $G$ is a
constant. This qualitatively explains the Tully-Fisher relation.  
The coefficient $\alpha$ of the Tully-Fisher relation can be obtained by expanding 
$b(\sqrt{GM})$: $b=b_{0}+b_{1}\sqrt{GM}+b_{2}GM+...$. Without field self-coupling 
(i.e. setting $\sqrt{GM}=0$ in Eq.~\ref{eq:Lagrangian}), $b=0$ so $b_{0}=0$. 
Since $\sqrt{GM}\gg GM$ we have at leading order $ln(M)=4ln(v)+ln(2\pi \tau/\sqrt{G}b_{1})$. 
We remark that the Tully-Fisher relation is not explained in the dark matter
scenario, and is a built-in feature of MOND as are the flat rotation
curves. 

Dark matter was first hypothesized to reconcile the motions of galaxies
inside clusters with the observed luminous masses of those clusters.
Estimating the non-Abelian effects in galaxy clusters with our 
technique is difficult: 1) the force \emph{outside} the galaxy is 
suppressed since the binding of the galaxy 
components increases (this will be discuss further at the end
of the Letter), but 2) the non-Abelian effects on the remaining outside 
field could balance this if the remaining outside field is strong enough. 
Since clusters are made mostly of elliptical galaxies 
for which the approximate sphericity suppresses the non-Abelian effects
inside them, we ignore the first effect. We assume furthermore that the 
intergalactic gas is distributed homogeneously enough so that non-Abelian 
effects cancel (i.e. the gas does not influence our computation). Finally, we
restrict the calculation to the interaction of two galaxies, assuming
that others do not affect them. With these three assumptions,
we can apply our calculations. Taking 1 Mpc as the distance between the two
galaxies and M=40$\times$10$^{9}$ M$_{\odot}$ as the luminous mass of
the two galaxies, we obtain $b=-0.012$ in lattice units.
We express this from the dark matter standpoint by forcing gravity 
to obey a Newtonian form:
\begin{eqnarray}
V(r)=-G\frac{M}{2}(\frac{1}{r}-\frac{b}{a}r) & 
\equiv & -G\frac{M'}{2}\frac{1}{r}
\label{eq:pot cluster}
\end{eqnarray} 
\noindent with $M'/M=1-r^{2}b/a=251.$ Gaseous mass in a cluster is 
typically 7 times larger than the total galaxy mass. Assuming that 
half of the cluster galaxies are spirals or flat ellipticals for which the
non-Abelian effects on the remaining field are neglected, we obtain 
for the cluster a ratio $(M'/M)_{cluster}=18.0$, that is our model of 
cluster is composed of 94\% dark mass, to be compared with the 
observed 80-95\%. 

Non-Abelian effects emerge in asymmetric
mass distributions. This makes our mechanism naturally compatible with the 
Bullet cluster observation~\cite{bullet and DM} (presented as a direct proof of 
dark matter existence since it is difficult
to interpret in terms of modified gravity): Large non-Abelian effects 
should not be present in the center of the cluster collision where the 
intergalactic gas of the two 
clusters resides if the gas is homogeneous and does not show large asymmetric 
distributions. However, the large non-Abelian effects discussed in the preceding
paragraph still accompany the galaxy systems. 

In addition to reproducing the rotation curves and cluster dynamics and to explain
the Tully-Fisher relation,
our approach implies several consequences that can be tested: 
\textbf{1)} Since the Non-Abelian distortions of the field are suppressed for 
spherically homogeneous distributions, rotation curves closer to Newtonian
curves should be measured for spherical galaxies; \textbf{2)} Two spiral 
galaxies should interact less than a similar system formed by two 
spherical galaxies. \textbf{3)}
In a two-body system, we expect a roughly linear potential for large
enough effective coupling  ($\gtrsim10^{-3}$). This may be testable
in a sparse galaxy cluster; \textbf{4)} The past universe
being more homogeneous, and density fluctuations being less massive,
the non-Abelian effects should
disappear at a time when the universe was homogeneous enough; 
\textbf{5)} Structure formations would proceed differently than presently thought
since dark matter is an ingredient of the current models, and since those
assume an Abelian potential. Particularly, models of mergers of galaxies
using a linear potential rather than dark matter constitute
another test. 

Although the consequences of non-Abelian effects in gravity for galaxies
are not familiar, similar observations (increases of a force's 
strength at large distance) are well known in sub-nuclear physics. Those, 
closely related to the confinement of non-relativistic quarks inside hadrons, 
are fully explained 
by the theory of the strong nuclear force (Quantum Chromodynamics, QCD). 
First, QCD is the archetype non-Abelian theory. Second, although
the quark color charge is only unity, the QCD effective coupling
$\alpha_{s}^{eff}$ is large at the scale of the nucleon size (about 1 at
$10^{-15}$ m~\cite{alpha_s}) so the overall force's intensity is large, as for
massive systems in gravity. These are the two ingredients needed to 
confine quarks: The gluons emitted by the non-relativistic quarks
strongly interact with each other and collapse into string-like flux tubes. 
Those make the strong force to be constant for distances $r\simeq10^{-15}$ m
rather than displaying an $\alpha_{s}^{eff}(r)/r^{2}$ dependence,
see e.g.~\cite{QCD}. We also remark that a relation akin to the Tully-Fisher one
exists for the strong force in the confinement regime: the angular momenta
and squared masses of hadrons are linearly correlated. These ``Regge trajectories''
are at the origin of the string picture of the strong force. Lattice techniques are 
a well developed tool to study gluon-gluon interactions at large 
distances. Hence, it was a ready-to use tool for our purpose.
The simplest lattice QCD calculations displaying quark confinement 
are done in the {}``gluonic sector'', that is without dynamical 
quark degrees of freedom). Similarly, our calculation excluded $T^{\mu \nu}$, 
the sources of $\psi$ in the Lagrangian $\mathcal{L}$. We also note that
the QCD Lagrangian has a similar structure as $\mathcal{L}$ in 
Eqs.~\ref{eq:Einstein-Hilbert} and~\ref{eq:Lagrangian}. The close analogy 
between gravity and QCD is the reason we used the particle physics terminology 
in this Letter. This analogy has been already noticed and discussed, see
e.g.~\cite{QCD-GR analogy}.

Before concluding, we exploit further the QCD-gravity 
analogy, now on a qualitative level.
The confinement of gluons inside a hadron not only changes the $1/r$
quark-quark potential into an $r$ potential, but also causes two hadrons 
to not interact through the strong force~\cite{residual} since there is no strong
force carriers outside the hadrons. Similarly, the increased binding 
inside a galaxy would weaken its interaction with outside bodies. 
Such reduction of the strength of gravity is opposite to what we 
would conclude by explaining galaxy rotation
curves with hallos of exotic dark matter or with gravity modifications, and may
be relevant to the fact that the universe expansion is accelerating rather
than decelerating. This
is currently explained by the repulsive action of a dark energy,
see e.g.~\cite{review dark energy}. However, if gravity is weakened,
the difference between the assumed Abelian force and the actual strength 
of the force would be seen as an additional repulsive effect. 
Such effect would not explain a net repulsion 
since it would at most suppress the force outside of the mass system 
(as for QCD). Thus, it would not be directly 
responsible for a net acceleration of the universe expansion. 
Nevertheless, it may reduce the need for dark energy. To sum up, 
the gravity/QCD parallel propounds that dark energy 
may be partly a consequence of energy conservation between the 
increased galaxy binding energy vs. the outside effective potential 
energy. This would implies a quantitative relation between dark energy 
and dark matter, which might explain naturally the
\emph{cosmic coincidence problem}~\cite{review dark energy}. 

To conclude, the graviton-graviton interaction suggests a mechanism 
to explain galaxy rotation curves and cluster dynamics. Calculations 
done within a weak field approximation agree well with observations
involving dark matter, without requiring arbitrary parameters or
exotic particles. The Tully-Fisher relation arises naturally in our framework. 
Our approach hints that dark energy could partly
be a consequence of energy conservation between the increased galaxy 
binding energy and the outside potential energy.

\textbf{Acknowledgments:} We thank P.-Y. Bertin, S. J. Brodsky, V.
Burkert, G. Cates, F.-X. Girod, B. Mecking, A. M. Sandorfi and 
Xiaochao Zheng for useful discussions. We are grateful to the reviewer
of this Letter for pointing out the Tully-Fisher relation.

\vskip .1truein

%%%%%%%%%%%%%%%%%%%%%%%%%%%%%%%%%%%%%%%%%%%%%%%%%%%%%%%%%%%%%%%%%%%%%%%


\begin{references}
\bibitem{higher order}  The $\psi^{\mu \nu} \mid_{\mu\neq 0\nu\neq 0}$ terms that
are neglected in the operations leading to $\partial\psi\partial\psi\rightarrow
\partial_{\lambda}\psi^{00}\partial^{\lambda}\psi_{00}$ 
can be ignored in front of the ${\psi^{00}}\partial_{\lambda}\psi^{00}\partial^{\lambda}\psi^{00}$ 
term: The $\partial\psi\partial\psi$ do not contribute to the field self-interaction so they
are irrelevant to the effects we are studying. $\partial_{\lambda}\psi^{00}\partial^{\lambda}\psi_{00}$  dominates these other
$\partial_{\lambda}\psi^{\mu \nu}\partial^{\lambda}\psi_{\mu \nu}$. In our results, 
these should be a correction to a 
normalization factor ($a$). Because of the simplicity of our galaxy and cluster
models, we ignore such corrections. Likewise, the corrections coming
from the modification of the Euler-Lagrange equation of motion enter only at the level of
the terms $\psi^{n+1}\partial^{2}\psi$, with $n \geq 2$.
\bibitem{m+m} The additive form of the global coupling is supported by the $G^{2}$ 
correction to the precession of the 
perihelion of the orbit of two bodies of masses $m_{1}$ and $m_{2}$, 
that is given by $\frac{Gm_{1}m_{2}}{r}(\frac{G(m_{1}+m_{2})}{2r})$, or in
dominant one-loop correction to the Newtonian potential established in quantum 
gravity (i.e. the first order graviton self-interaction): 
$V(r)=\frac{Gm_{1}m_{2}}{r}(1+\frac{3G(m_{1}+m_{2})}{2r})$~\cite{Kiefer}. Comparison
of these corrections to our global coupling are relevant because they stem in 
part from the field-self interaction.
Clearly, our estimate of the magnitude of the global coupling is naive. In particular,
the use of the field superposition principle is inadequate when large non-linear
effects are present.
\bibitem{factorials} The factors 3! and 4! cancel the combinatorial factors 
that arise during the calculation of the green function when a Wick contraction 
is done. Hence, each Feynman graph associated with a 
$\phi^{n}\partial\phi\partial\phi$ term has a coupling $G^{(n-2)/2}$ at an $n-$legs vertex
rather than ${n!G}^{(n-2)/2}$.
\end{references}
\end{document}